\documentclass[twocolumn]{revtex4}
\usepackage{graphicx,epsfig}
\usepackage{amsmath}
\usepackage{amsfonts}
\usepackage{xcolor}
\usepackage{fancyhdr}
\usepackage{hyperref}
\DeclareUnicodeCharacter{2212}{-}

\begin{document}




\title{Eigenvector Localization and Universal Regime Transitions in Multiplex Networks: A Perturbative Approach}
\author{Joan Hernàndez Tey}
\affiliation{Facultat de F\'{\i}sica, Universitat de Barcelona, 08028 Barcelona, Spain.}

\author{Emanuele Cozzo}
\affiliation{Departament de F\'isica de la Mat\`eria Condensada, Facultat de F\'{\i}sica, Universitat de Barcelona, 08028 Barcelona, Spain.\\ Universitat de Barcelona Institute of Complex Systems (UBICS), Universitat de Barcelona, 08028 Barcelona, Spain. \\ Communication Networks \& Social Change research group, Universitat Oberta de Catalunya, Barcelona, Spain.}

\begin{abstract}
We study the transition between layer-localized and delocalized regimes in a general contact-based contagion model on multiplex networks. Using the inverse participation ratio, we characterize how activity shifts from being confined to a single layer to spreading across the entire system. Through a first-order perturbative analysis of the leading eigenvector of the supra-contact probability matrix, we derive an analytical expression for the fictive coupling $p^*$ that marks the crossover between the two regimes. This result reproduces and explains previously observed numerical scalings and extends them to a broad class of contact-based processes beyond the Susceptible–Infected–Susceptible model. We also obtain an analytical expression for the IPR of the non-dominant layer in the localized regime, confirming its power-law dependence on the coupling with exponent $\alpha=4$. Finally, we study the transition between non-dominant and dominant layers as a function of the intra-layer activity parameter $\gamma$. Our analytical findings are supported by dynamical simulations that highlight distinct susceptibility patterns across regimes. Altogether, this work provides a unified spectral framework for understanding localization and dominance transitions in multiplex contagion dynamics.
\end{abstract}

\maketitle


\section{Introduction}
 Understanding how activity localizes in complex systems, from disease spreading to information diffusion, is a central problem in network science. In particular, the transition between localized and delocalized regimes, where activity either remains confined to specific parts of the system or spreads broadly, has been studied extensively in heterogeneous monoplex networks. Beyond their theoretical interest, these transitions are relevant for practical applications such as controlling epidemic outbreaks or the spread of information.

 Since the inception of multiplex network theory \cite{cozzo2018multiplex,kivela2014multilayer}, considerable effort has been devoted to identifying how critical behaviors in these systems differ from their monoplex counterparts. A central focus of this research has been to relate these behaviors to the spectral properties of matrices that govern the dynamics on multiplex networks, such as the supra-adjacency or supra-Laplacian matrices \cite{sanchez2014dimensionality}. A phenomenon that is intrinsically tied to the multiplex nature of these systems is layer-localization.

 Localization phenomena are strictly related to a phase transition between an active and an inactive state \cite{silva2021dissecting}. In monoplex heterogeneous networks, localization has been extensively studied in the context of propagation models for spreading of rumors and diseases. It refers to a metastable state in which activity remains confined to a small, non-extensive subset of nodes, typically those associated with hubs or dense subgraphs. This behavior is generally attributed to quenched disorder, i.e., structural heterogeneity, in the network \cite{goltsev2012localization, ferreira2016metastable, pastor2018eigenvector}.

 However, multiplex networks give rise to a different form of localization. In this case, activity can be localized not only on a few nodes but on entire layers of the network, even when each individual layer is structurally homogeneous. This layer-localized to delocalized transition was first reported in the context of the Susceptible-Infected-Susceptible (SIS) model in \cite{de_Arruda_2017}. Remarkably, the active state of the system can become confined to an entire layer, making localization an extensive phenomenon that occurs in the absence of intra-layer disorder.

 As in the case of monoplex networks, this behavior can be understood by analyzing the spectral properties of the matrix that governs the dynamics; in this case, the supra-adjacency matrix of the multiplex network or some function of it. Specifically, since the steady-state infection probabilities can be expressed as a linear combination of the eigenvectors of the dynamical operator, the emergence of localized states is tied to the localization of the leading eigenvector \cite{goltsev2012localization, Cozzo_2013, de_Arruda_2017}.

 To illustrate this, let us consider the case of a two-layer multiplex network. The corresponding eigenvalue problem for the supra-adjacency matrix reads:

\begin{equation}
\left[ \begin{array}{c|c}
A_1 & pI\\ \hline
pI & A_2 \\ 
\end{array}
\right] \cdot \left[ \begin{array}{c} 
v_1 \\ \hline
v_2 \\ 
\end{array} \right] = \lambda \cdot 
\left[ \begin{array}{c} 
v_1 \\ \hline
v_2 \\ 
\end{array} \right]
\label{supraA}
\end{equation}

where
$A_1$ and $A_2$ are the adjacency matrices of each layer, $p$ is the interlayer coupling strength, and $v_1$, and $v_2$ are the components of the eigenvector associated with the respective layers.

As shown in \cite{Cozzo_2013}, depending on the value of $p$, the leading eigenvector can be either localized in one layer - the dominant one, i.e., the layer with the largest leading eigenvalue — or delocalized across all layers. In the localized regime, only the components of the eigenvector associated with the dominant layer are of order $O(1)$, while the rest are negligible. In the delocalized regime, all components are of the same order.

A commonly used metric to quantify localization is the Inverse Participation Ratio (IPR) \cite{silva2021dissecting}, defined for a normalized eigenvector $v$ as:

\begin{equation}
IPR(\mathbf{v})=\sum_{i=1}^N v_i^4,
\label{IPR}
\end{equation}

where $N$ is the number of nodes. In this context, the IPR distinguishes between different localization regimes: it remains finite when activity is localized on a small subset of nodes (non-extensive localization), scales as 
$N^{−\nu}$ with $0<\nu<1$ for subextensive localization, and scales as 
$N^{-1}$ when the activity is delocalized across the network\cite{silva2021dissecting}. This classification has been used, for example, to show that the principal eigenvector of networks with power-law degree distributions can be localized depending on the value of the degree exponent \cite{goltsev2012localization, silva2021dissecting}.

In multiplex networks, the IPR method can be similarly applied to the leading eigenvector of the supra-adjacency matrix. For a two-layer system, the IPR of the global eigenvector can be decomposed as $IPR(v)=IPR(v_1)+IPR(v_2)=IPR_1+IPR_2$, where $IPR_\alpha$ refers to the contribution from layer $\alpha$. This decomposition allows us to track how localization shifts between layers. From a dynamical perspective, the modified susceptibility also reflects this transition: in the localized regime, it exhibits a double peak with only one diverging; in the delocalized regime, a single diverging peak appears \cite{de_Arruda_2017}.

More recently, Ferraz de Arruda et al.~\cite{Ferraz_de_Arruda_2020} showed that the IPR curves for different systems collapse onto a universal curve when the coupling parameter $p$ is rescaled by a characteristic value $p^*$, which they show depends linearly on the difference in mean degree between layers: $p^* \approx \beta_1 \langle k_1 - k_2 \rangle + \beta_2$, with $\beta_1$ and $\beta_2$ constants. This characteristic value marks the crossover between the layer-localized and delocalized regimes and is determined by identifying the point at which the power-law decay of the IPR's contribution of the non-dominant layer $\text{IPR}_2$ in the localized regime, characterized by the relation $\log(\text{IPR}_2) \approx \alpha \log(p) + c$ with $\alpha \approx 4$ and $c$ depending on network structure, intersects with the plateau typical of the delocalized regime. By analogy to the fictive temperature defined in glassy systems~\cite{mauro2009fictive}, we refer to this transition point as the \textit{fictive coupling}.

In this work, we provide an analytical argument that explains the numerical results reported in~\cite{Ferraz_de_Arruda_2020}, showing that the fictive coupling $p^*$ separating the layer-localized and delocalized regimes can be derived through a perturbative analysis of the leading eigenvector of the supra-contact probability matrix. This approach not only clarifies the origin of the scaling observed numerically, but also allows us to generalize the result beyond the SIS model to a broader class of contact-based contagion processes. Furthermore, we establish a connection between this phenomenon and another critical behavior previously studied in~\cite{Cozzo_2013}, namely, the switching of the dominant layer as a function of intra-layer activity.

\section{Supra contact probability matrix}
Let us now define the main mathematical object of our analysis.The contact probability matrix $\mathbf{R(A)}=\{R(A)_{uv}\}$ was introduced in \cite{gomez2010discrete} in the context of a contact-based Markov chain approach to epidemic spreading in single-layer networks. Its elements are given by
\begin{equation}
R(A)_{uv}=1-\left(1-\frac{A_{uv}}{k_{u}}\right)^{\gamma_{u}},
\end{equation}
with $A_{uv}$ the elements of the adjacency matrix $\mathbf{A}=\{A_{uv}\}$, and $k_u$ the degree of the node $u$. The element $R(A)_{uv}$ is the probability of node $u$ contacting node $v$ and depends on the activity parameter $1\leq \gamma_u$. When $\gamma_u=\gamma=1$, $R(A)_{uv}=\frac{A_{uv}}{k_u}$ and it describes a contact process, conversely, when $\gamma\to\infty$, $R(A)_{uv}\to A_{uv}$, and the fully reactive SIS process is recovered.
In \cite{Cozzo_2013}, the contact probability matrix is generalize to a multiplex network. 
Consider a multiplex system composed of $M$ layers with the same $N$ nodes in each. Let $u,v \in [1, N]$ be node indices in a given layer, while Greek letters such as $\alpha, \kappa \in [1, M]$ characterize the layer index. To refer to nodes across the entire multiplex system, global indices $i, j \in [1, MN]$ are introduced. Thus, a bijective mapping can be defined between the two sets of indices given by $i=u\alpha$.
In this scenario, we have $R_{\alpha}=R(A_\alpha)$ for each layer. Thus, the supra-contact probability matrix is given by
\begin{equation}
\bar{R}=\bigoplus_{\alpha} R_{\alpha}+ pC,
\label{barR}
\end{equation}
where the first part is the direct sum of each layer contact probability matrix and C is a matrix where $C_{ij}=1$ if $i$ and $j$ are global indices associated to the same node in different layers. Therefore, the supra-contact probability matrix has a block structure with $R_{\alpha}$ on the diagonal and the $p*C$ out of this diagonal, for instance
$
\bar{R}=\left( \begin{array}{c|c} 
R_{1} & C_{12} \\ \hline
C_{21} & R_{2}\\ 
\end{array}\right)
$ for a two layer multiplex network.
Furthermore, when considering the discrete-time evolution equation for the contagion probability of a node $i$ \cite{Cozzo_2013}, we obtain
\begin{align}
\vec{p}(t+1)=&[\vec{1}-\vec{p}(t)]*[\vec{1}-\vec{q}(t)]+(1-\mu)*\vec{p}(t)
\label{prob}
\\ \nonumber &+\mu*[\vec{1}-\vec{q}(t)]*\vec{p}(t),
\end{align}
where $q_i(t)=\prod_j[1-\beta R_{ij}p_j(t)]$ is the probability to not be infected for the node $i$, with $p=\frac{\epsilon}{\beta}$, where $\beta$ characterizes the contagion rate within layers and $\epsilon$ characterizes the contagion rate between the same node in different layers.$\mu$ being the probability of an active state decaying to an inactive one. The phase transition to an endemic phase is given by$\left(\frac{\beta}{\mu}\right)_c=\frac{1}{\Lambda_{max}}$ \cite{Cozzo_2013}, where $\Lambda_{max}$ is the largest eigenvalue of the supra-contact probability matrix \eqref{barR}.

\section{Perturbative analysis}

In this section, we study the localization properties of the leading eigenvector of the supra-contact probability matrix through a perturbative analysis of the IPR to obtain the characteristic value of the localized-delocalized regime transition, i.e. the fictive coupling $p^*$. We first consider random regular networks (RRNs) due to the simplicity of the calculations. Later, we will show that these results are generalizable to other networks models. 

\begin{figure}
\centering
\centering
\includegraphics[width=1\linewidth]{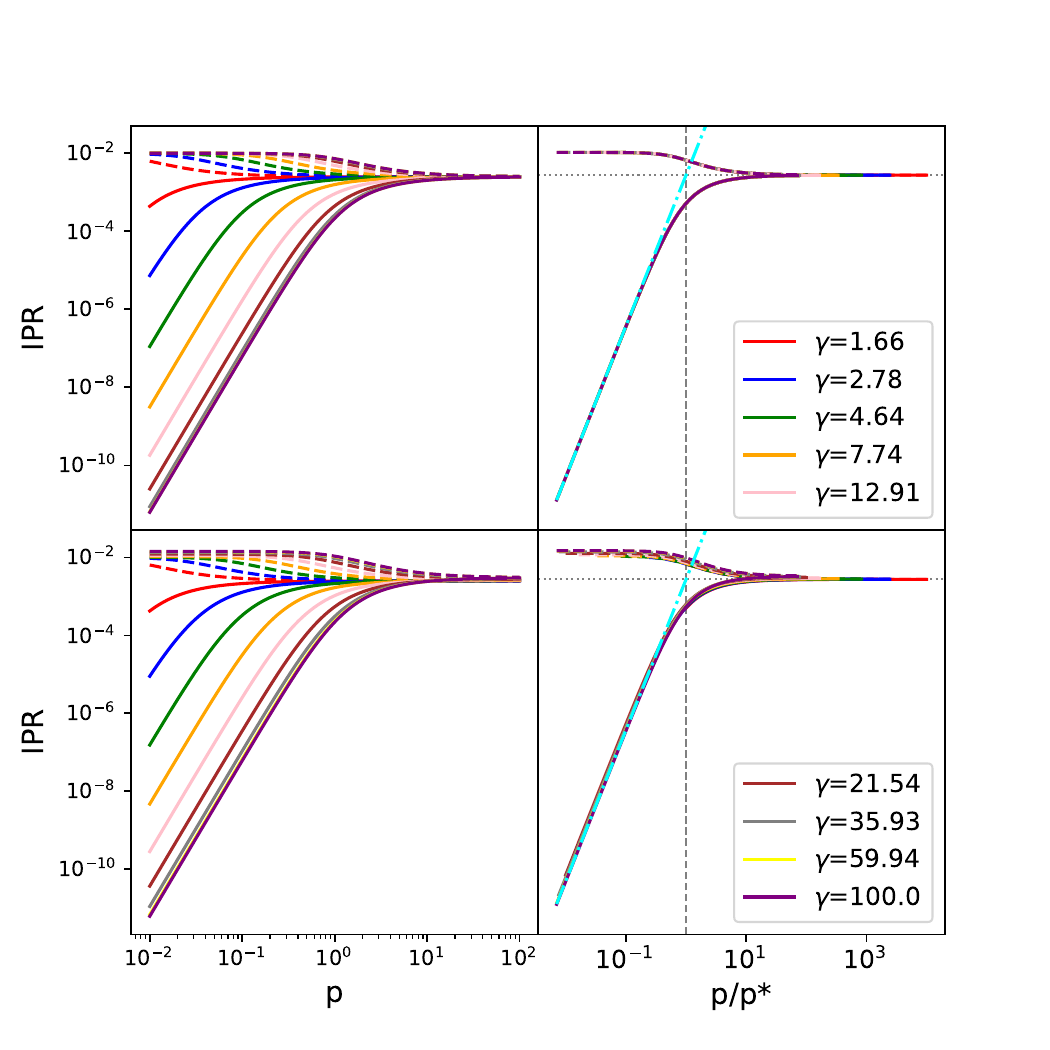}
\caption{Inverse participation ratio (IPR) for the dominant (dotted line) and non-dominant (solid line) layers as a function of the coupling $p$ (left) and the rescaled coupling $p/p^*$ (right). Top panels: random regular networks with $k_1 = 10$, $k_2 = 8$, and 100 nodes. Bottom panels: Erdős--Rényi networks with $\langle k_1 \rangle = 10$, $\langle k_2 \rangle = 8$, and 1000 nodes. The blue line corresponds to the perturbative result given by Eq.~(\ref{Ipr2approx})}
\label{fig:1}
\end{figure}

Let us consider the case where $p\ll1$. In the supra-contact probability matrix $\bar{R}=\bigoplus_{\alpha}R_{\alpha}+pC$, the p-dependent part can be treated as a perturbation. Therefore, we can approximate the leading eigenvalue $\bar{\Lambda} \approx \Lambda_0 + p\Delta \Lambda $ and the corresponding leading eigenvector $\vec{v} \approx \vec{v}_0 + p\Delta \vec{v}$, where $\Lambda_0$ and $\vec{v}_0$ are the leading eigenvalue and eigenvector of the uncoupled system. For the uncoupled system, $R=\bigoplus_{\alpha}R_{\alpha}$ is a diagonal block matrix, its eigenvalues are those of the $R_{\alpha}$, thus $\Lambda_0$ is the leading eigenvalue of the dominant layer and $\vec{v}_0=\left[ \begin{array}{c} 
v_0 \\ \hline
0 \\ 
\end{array} \right]$ is its associated eigenvector, with $v_0$ the leading eigenvector of the dominant layer. Considering RRNs, the largest eigenvalue of each layer is $\Lambda_{\alpha}=k_{\alpha}\cdot(1-(1-\frac{1}{k_{\alpha}})^{\gamma_{\alpha}})$ and its corresponding eigenvector is a normalized vector of all ones.  The first-order approximation is given by (assuming normalized eigenvectors)
\begin{equation}
\Delta \Lambda_{max}= \vec{v}^{\top}_0 C \vec{v}_0
\end{equation}
\begin{equation}
\Delta \vec{v} = \sum_{u_0 \not= v_0}\frac{\vec{u}^{\top}_0 C \vec{v}_0}{\Lambda_{v_0}-\Lambda_{u_0}}\vec{u}_0,
\label{sum}
\end{equation}
where $\vec{u}_0$ are the non-leading eigenvectors.
Since the leading eigenvectors of two RRNs are equal, the non-leading eigenvector of one of the layer are orthogonal also to the leading eigenvector of the other, there is just one term in the sum in \eqref{sum} that is not equalto zero. Therefore, $\Delta \Lambda=0$ and $\Delta \vec{v}=\frac{p}{\Lambda_1-\Lambda_2}C\vec{v}_0$. Consequently, the resulting vector is 
\begin{equation}
\vec{v}=\left( \begin{array}{c} 
\vec{v}_0 \\ \hline
\frac{p}{\Lambda_1-\Lambda_2}\vec{v}_0 \\ 
\end{array}\right).
\label{leadingv}
\end{equation}

Now, we can obtain the fictive coupling $p^*$ as $p$ at which $IPR_1\approx IPR_2$ as defined in \cite{Ferraz_de_Arruda_2020}, therefore, 
\begin{equation}
p^*\approx \Lambda_1- \Lambda_2,    
\label{pstar}
\end{equation}
that for RRNs reads:
\begin{equation}
p^*\approx k_1\left(1-\left(1-\frac{1}{k_1}\right)^{\gamma_1}\right)-k_2\left(1-\left(1-\frac{1}{k_2}\right)^{\gamma_2}\right).
\end{equation}
It is worth noting that our perturbative expansion relies on the difference between the leading eigenvalues of the two uncoupled layers. A degeneracy, i.e., $\Lambda_1=\Lambda_2$, would correspond no dominant layer. In such a case, the concept of a layer-localized regime loses meaning, and therefore no layer-delocalization transition can occur, that is equivalent to say $p^*=0$.

It results that when $p$ is scaled with $p*$, the IPR curves for different $\gamma$ values collapse onto a universal curve, as shown in  Fig.~\ref{fig:1}, generalizing the result of \cite{Ferraz_de_Arruda_2020}. 
These results have been obtained assuming that we have a RRN in each layer, thus reducing to only one non-zero term the summation in (\ref{sum}). However, the summation of the products between non-dominant layer non-leading eigenvectors and the leading eigenvector of the dominant layer vanishes on average for other network models as well, leading to the same result (see Fig.~\ref{fig:1} bottom panel).

Considering the $IPR_2$ for $p<p^*$, from the perturbative results \ref{leadingv} we have
\begin{equation}
\label{Ipr2approx}
IPR_2\approx \frac{1}{4N}\left(\frac{p}{\Lambda_1-\Lambda_2}\right)^4=\frac{1}{4N} {p^*}^4, 
\end{equation}
where the factor $\frac{1}{4N}$ comes from the normalization. 

The numerical result given in \cite{Ferraz_de_Arruda_2020} is $\log{IPR_2}=\alpha\log(p)+c_1$. They obtained that $\alpha\approx4$ and $c_1$ was dependent on the network structure and size, and it coincides with our analytical result $\alpha=4$ and $c_1 = -4\log(\Lambda_1-\Lambda_2)-\log 4N$.

As we mentioned in the introduction section, another numerical results showed in \cite{Ferraz_de_Arruda_2020}, was the linear relation between the fictive coupling $p^*$ and the difference between the average degree of the layers, expressed as $p^*=\beta_1 < k_1 - k_2 > + \beta_2$.

In particular, the numerical value reported in \cite{Ferraz_de_Arruda_2020} are $\beta_1\approx 1.2$ and $\beta_2\approx 0.4$, obtained by fitting values of $p^*$ for different multiplex networks composed by two layers, with Erdös-Rényi networks in both layers, or  Erdös-Rényi networks in one layer, and power-law networks in the other. This linear dependence is a direct consequence of the perturbative result \ref{pstar} that is a linear dependence of $p^*$ on the difference of the leading eigenvalues of the layers, and the relation between the leading eigenvalue of a homogeneous network and its average degree. In particular, for multiplex networks composed of RRNs in both layers $\beta_2$ should be strictly zero and $\beta_2\approx 0$ for multiplex networks with Erdös-Rényi layers. The nonvanishing value of $\beta_2$ obtained in \cite{Ferraz_de_Arruda_2020} is thus due to the presence of non homogeneous layers, for which the approximation $\Lambda_\alpha \approx <k_\alpha>$ is not true. 

\section{Non-dominant to dominant transition}

The non-dominant to dominant transition occurs when $\gamma$ changes for one of the non-dominant layers, as $\gamma$ increases, the leading eigenvalue associated with the layer eventually crosses the value of the dominant layer becoming dominant \cite{Cozzo_2013}. Since, as we showed in the previous section, the layer-localized to delocalized transition can be described interely by perturbative analysis of those eigenvalues, it is interesting to study how IPR behaves through this transition.

Let us consider a system of two RRNs with the first layer with $\gamma_1\xrightarrow{}\infty$ and $k_1$ and the second with $k_2 > k_1$ and a variable $\gamma$. In this scenario, for low enough values of $\gamma$, the dominant layer is the first as it has the larger eigenvalue. Then, when surpassing $\gamma*$ (the point where $\Lambda_1=\Lambda_2$), there is a dominance change where the second layer has the largest eigenvalue. As we know the expression for the eigenvalue in a RRN, we can calculate the analytical expression of $\gamma*$

\begin{equation}
\gamma*=\frac{\ln{\left(1-\frac{\Lambda_1}{k_2}\right)}}{\ln{\left(1-\frac{1}{k_2}\right)}}
\end{equation}

This result is exact for a RR network. For an Erdös-Rényi network, we can approximate $\gamma*$ using $\Lambda\approx <k>+1$. 

\begin{figure}
\centering
\includegraphics[width=0.9\linewidth]{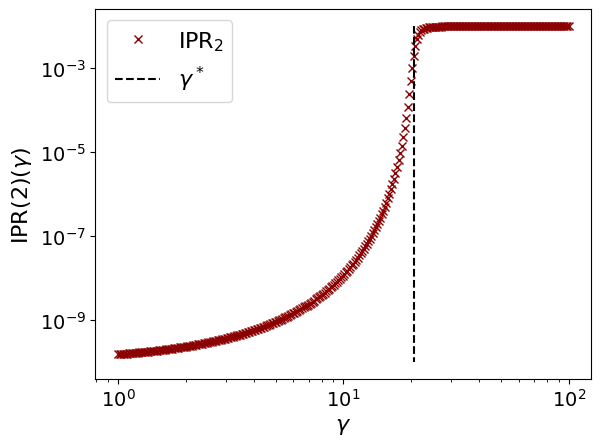}
\caption{Representation of the inverse participation ratio $\mathrm{IPR}_2$ as a function of $\gamma$ for a two-layer multiplex network with fixed coupling $p = 0.1$. One layer has $\gamma_1 \to \infty$, while the other has variable $\gamma_2 = \gamma$. Both layers are random regular networks with 100 nodes, $k_1 = 10$, and $k_2 = 12$. The vertical dotted line indicates the transition point at $\gamma^* = 20.59$.}
\label{fig:5}
\end{figure}

The change of dominance is shown in Fig.~\ref{fig:5}, where the IPR for the second layer for low values of $\gamma$ has a nondominant form and, then, when $\gamma^*$ is surpassed, it behaves as the dominant layer. Thus, the IPR of the non-dominant layer can be used as an order parameter for this transition.

\section{Dynamical behaviour at the regimes transitions}

\begin{figure}[t]
\centering
\includegraphics[width=\columnwidth]{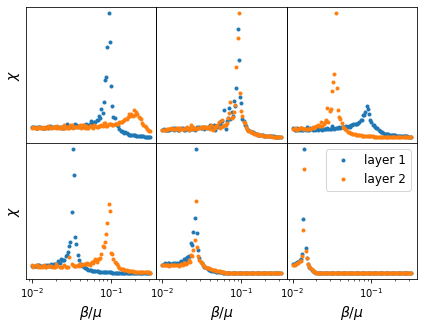}
\caption{Top panels: modified susceptibility $\chi$ as a function of $\beta/\mu$ for $\eta = 0.1$. Layer~1 has $k_1 = 10$, $\Lambda_1 \approx 11$, and $\gamma_1 \to \infty$; layer~2 has $k_2 = 30$, $\Lambda_2 \approx 31$, and variable $\gamma_2$. From left to right: (i) dominance of layer~1 ($\gamma_2 = 5$); (ii) near the dominance transition ($\gamma_2 = 13.5$); (iii) dominance of layer~2 ($\gamma_2 = 100$). The transition occurs at $\gamma^* = 13.47$. \\ Bottom panels: modified susceptibility $\chi$ as a function of $\beta/\mu$ for different values of the coupling $\eta$. Layer~1 has $k_1 = 30$, $\Lambda_1 \approx 31$, and $\gamma_1 \to \infty$; layer~2 has $k_2 = 10$, $\Lambda_2 \approx 11$, and $\gamma_2 = 10^4$. From left to right: (iv) localized regime ($\eta = 0.01$); (v) near the localization transition ($\eta = 15$); (vi) delocalized regime ($\eta = 50$). The fictive coupling is $p^* = 20$.
}
\label{fig:sample}
\end{figure}

We perform Monte Carlo simulations on an ER two-layer multiplex network, utilizing a quasi-stationary algorithm \cite{deoliveira2004simulate} to simulate the contagion process. Our focus is on examining the differences in the dynamical behavior associated with the two regime transitions discussed earlier: the layer-localized to delocalized transition and the dominance transition. These transitions can be characterized using the modified susceptibility, defined as
\begin{equation}
\chi=N\frac{\langle \rho \rangle^2-\langle \rho^2 \rangle}{\langle \rho \rangle},
\end{equation}

\noindent
where $\rho$ is the density when the stationary state is reached and $N=nM$. The susceptibility as a function of the structural parameter exhibits a diverging peak at the point of transition, with the peak position being inversely related to the largest eigenvalue of the system. Importantly, since we are concerned only with the peak positions, the normalization factor is not of interest. In a multiplex network, we can analyze the density of each layer individually. In this context, each layer's susceptibility shows a peak; however, only the peak corresponding to the dominant layer will diverge if the system is in the layer-localized regime, or both diverges when the system is delocalized.

We conducted simulations by varying network structural parameters to study the transitions analyzed in previous sections. To investigate the localized-delocalized transition, we varied $\eta=\frac{\epsilon}{\beta}$ which corresponds to the coupling parameter $p$ in the IPR study. For the dominant-to-non-dominant transition, we varied  $\gamma_2$ for the second layer.

In Fig. \ref{fig:sample}, we illustrate different structural configurations that are informative for studying these regimes. The top panels focus on the dominance transition. When $\gamma_2^*>\gamma_2$ , layer one dominates, with a diverging peak near $\frac{\beta}{\mu}=\frac{1}{\Lambda_1}$ , while layer two shows its peak near $\frac{\beta}{\mu}=\frac{1}{\Lambda_2}$. As $\gamma_2$ increases, $\Lambda_2$ also increases, causing the peak to shift. When $\Lambda_1=\Lambda_2$, both layers exhibit a diverging peak at the same position and when $\Lambda_1<\Lambda_2$ the second layer's peak diverges near $\frac{\beta^*}{\mu}=\frac{1}{\Lambda_2}$. The bottom panels focus on the localized to delocalized transition. Layer one consistently shows a diverging peak near $\frac{\beta}{\mu}=\frac{1}{\Lambda_1}$. . For low values of $\eta$ the second layer exhibits a peak near $\frac{\beta}{\mu}=\frac{1}{\Lambda_2}$, however, as $\eta$ increases, this secondary overlap with that of the dominant layer. When $\eta\gg\eta^*$, layers one and two behave similarly, both showing a shifted diverging peak.

\section{Conclusions}

In this work, we investigated the transition between layer-localized and delocalized regimes in a general contact-based contagion model on multiplex networks. Using a first-order perturbative analysis of the leading eigenvector of the supra-contact probability matrix, we derived an analytical expression for the \textit{fictive coupling} $p^*$ that separates these regimes. This result not only reproduces the numerical scaling found in \cite{Ferraz_de_Arruda_2020}, but also provides a theoretical foundation for its universality, extending its validity beyond the SIS model to a broader class of contact-based dynamics.

We further obtained an analytical expression for the inverse participation ratio (IPR) of the non-dominant layer within the localized regime, confirming the power-law scaling with exponent $\alpha=4$ observed numerically. This scaling collapse confirms that the localized–delocalized crossover follows a universal functional form governed by the difference in layer spectral properties. This demonstrates that the observed universal behavior of the IPR has a spectral origin rooted in the perturbative structure of the supra-contact matrix.

In addition, we characterized the transition between dominant and non-dominant layers as a function of the intra-layer contact parameter $\gamma$. These results clarify the interplay between intra-layer activity and inter-layer dominance.

Finally, our numerical simulations confirmed the analytical results and revealed distinct dynamical signatures for the two transitions: the layer-localized/delocalized crossover and the dominance transition. The divergence patterns of the susceptibility peaks offer a clear phenomenological distinction between them.

Altogether, this work provides a unified spectral framework to understand localization and dominance transitions in multiplex contagion dynamics. While our perturbative approach captures the smooth nature of the observed crossover, a natural direction for future research is to explore whether, in suitable limits, the transition can become sharp, potentially associated with eigenvalue crossings or spectral gaps, typically associated with structural transitions in multiplex
networks.

\begin{acknowledgments}
We acknowledge financial supportsupport from the Spanish grant PID2021-128005NB-C22, and from Generalitat de Catalunya underproject 2021-SGR-00856.
\end{acknowledgments}

\bibliographystyle{apsrev4-2}
\bibliography{Ref.bib}

\begin{thebibliography}{13}%
\makeatletter
\providecommand \@ifxundefined [1]{%
 \@ifx{#1\undefined}
}%
\providecommand \@ifnum [1]{%
 \ifnum #1\expandafter \@firstoftwo
 \else \expandafter \@secondoftwo
 \fi
}%
\providecommand \@ifx [1]{%
 \ifx #1\expandafter \@firstoftwo
 \else \expandafter \@secondoftwo
 \fi
}%
\providecommand \natexlab [1]{#1}%
\providecommand \enquote  [1]{``#1''}%
\providecommand \bibnamefont  [1]{#1}%
\providecommand \bibfnamefont [1]{#1}%
\providecommand \citenamefont [1]{#1}%
\providecommand \href@noop [0]{\@secondoftwo}%
\providecommand \href [0]{\begingroup \@sanitize@url \@href}%
\providecommand \@href[1]{\@@startlink{#1}\@@href}%
\providecommand \@@href[1]{\endgroup#1\@@endlink}%
\providecommand \@sanitize@url [0]{\catcode `\\12\catcode `\$12\catcode `\&12\catcode `\#12\catcode `\^12\catcode `\_12\catcode `\%12\relax}%
\providecommand \@@startlink[1]{}%
\providecommand \@@endlink[0]{}%
\providecommand \url  [0]{\begingroup\@sanitize@url \@url }%
\providecommand \@url [1]{\endgroup\@href {#1}{\urlprefix }}%
\providecommand \urlprefix  [0]{URL }%
\providecommand \Eprint [0]{\href }%
\providecommand \doibase [0]{https://doi.org/}%
\providecommand \selectlanguage [0]{\@gobble}%
\providecommand \bibinfo  [0]{\@secondoftwo}%
\providecommand \bibfield  [0]{\@secondoftwo}%
\providecommand \translation [1]{[#1]}%
\providecommand \BibitemOpen [0]{}%
\providecommand \bibitemStop [0]{}%
\providecommand \bibitemNoStop [0]{.\EOS\space}%
\providecommand \EOS [0]{\spacefactor3000\relax}%
\providecommand \BibitemShut  [1]{\csname bibitem#1\endcsname}%
\let\auto@bib@innerbib\@empty
\bibitem [{\citenamefont {Cozzo}\ \emph {et~al.}(2018)\citenamefont {Cozzo}, \citenamefont {De~Arruda}, \citenamefont {Rodrigues},\ and\ \citenamefont {Moreno}}]{cozzo2018multiplex}%
  \BibitemOpen
  \bibfield  {author} {\bibinfo {author} {\bibfnamefont {E.}~\bibnamefont {Cozzo}}, \bibinfo {author} {\bibfnamefont {G.~F.}\ \bibnamefont {De~Arruda}}, \bibinfo {author} {\bibfnamefont {F.~A.}\ \bibnamefont {Rodrigues}},\ and\ \bibinfo {author} {\bibfnamefont {Y.}~\bibnamefont {Moreno}},\ }\href@noop {} {\emph {\bibinfo {title} {Multiplex networks: basic formalism and structural properties}}},\ Vol.~\bibinfo {volume} {10}\ (\bibinfo  {publisher} {Springer},\ \bibinfo {year} {2018})\BibitemShut {NoStop}%
\bibitem [{\citenamefont {Kivel{\"a}}\ \emph {et~al.}(2014)\citenamefont {Kivel{\"a}}, \citenamefont {Arenas}, \citenamefont {Barthelemy}, \citenamefont {Gleeson}, \citenamefont {Moreno},\ and\ \citenamefont {Porter}}]{kivela2014multilayer}%
  \BibitemOpen
  \bibfield  {author} {\bibinfo {author} {\bibfnamefont {M.}~\bibnamefont {Kivel{\"a}}}, \bibinfo {author} {\bibfnamefont {A.}~\bibnamefont {Arenas}}, \bibinfo {author} {\bibfnamefont {M.}~\bibnamefont {Barthelemy}}, \bibinfo {author} {\bibfnamefont {J.~P.}\ \bibnamefont {Gleeson}}, \bibinfo {author} {\bibfnamefont {Y.}~\bibnamefont {Moreno}},\ and\ \bibinfo {author} {\bibfnamefont {M.~A.}\ \bibnamefont {Porter}},\ }\href@noop {} {\bibfield  {journal} {\bibinfo  {journal} {Journal of complex networks}\ }\textbf {\bibinfo {volume} {2}},\ \bibinfo {pages} {203} (\bibinfo {year} {2014})}\BibitemShut {NoStop}%
\bibitem [{\citenamefont {S{\'a}nchez-Garc{\'\i}a}\ \emph {et~al.}(2014)\citenamefont {S{\'a}nchez-Garc{\'\i}a}, \citenamefont {Cozzo},\ and\ \citenamefont {Moreno}}]{sanchez2014dimensionality}%
  \BibitemOpen
  \bibfield  {author} {\bibinfo {author} {\bibfnamefont {R.~J.}\ \bibnamefont {S{\'a}nchez-Garc{\'\i}a}}, \bibinfo {author} {\bibfnamefont {E.}~\bibnamefont {Cozzo}},\ and\ \bibinfo {author} {\bibfnamefont {Y.}~\bibnamefont {Moreno}},\ }\href@noop {} {\bibfield  {journal} {\bibinfo  {journal} {Physical Review E}\ }\textbf {\bibinfo {volume} {89}},\ \bibinfo {pages} {052815} (\bibinfo {year} {2014})}\BibitemShut {NoStop}%
\bibitem [{\citenamefont {Silva}\ and\ \citenamefont {Ferreira}(2021)}]{silva2021dissecting}%
  \BibitemOpen
  \bibfield  {author} {\bibinfo {author} {\bibfnamefont {D.~H.}\ \bibnamefont {Silva}}\ and\ \bibinfo {author} {\bibfnamefont {S.~C.}\ \bibnamefont {Ferreira}},\ }\href@noop {} {\bibfield  {journal} {\bibinfo  {journal} {Journal of Physics: Complexity}\ }\textbf {\bibinfo {volume} {2}},\ \bibinfo {pages} {025011} (\bibinfo {year} {2021})}\BibitemShut {NoStop}%
\bibitem [{\citenamefont {Goltsev}\ \emph {et~al.}(2012)\citenamefont {Goltsev}, \citenamefont {Dorogovtsev}, \citenamefont {Oliveira},\ and\ \citenamefont {Mendes}}]{goltsev2012localization}%
  \BibitemOpen
  \bibfield  {author} {\bibinfo {author} {\bibfnamefont {A.~V.}\ \bibnamefont {Goltsev}}, \bibinfo {author} {\bibfnamefont {S.~N.}\ \bibnamefont {Dorogovtsev}}, \bibinfo {author} {\bibfnamefont {J.~G.}\ \bibnamefont {Oliveira}},\ and\ \bibinfo {author} {\bibfnamefont {J.~F.}\ \bibnamefont {Mendes}},\ }\href@noop {} {\bibfield  {journal} {\bibinfo  {journal} {Physical review letters}\ }\textbf {\bibinfo {volume} {109}},\ \bibinfo {pages} {128702} (\bibinfo {year} {2012})}\BibitemShut {NoStop}%
\bibitem [{\citenamefont {Ferreira}\ \emph {et~al.}(2016)\citenamefont {Ferreira}, \citenamefont {Da~Costa}, \citenamefont {Dorogovtsev},\ and\ \citenamefont {Mendes}}]{ferreira2016metastable}%
  \BibitemOpen
  \bibfield  {author} {\bibinfo {author} {\bibfnamefont {R.}~\bibnamefont {Ferreira}}, \bibinfo {author} {\bibfnamefont {R.}~\bibnamefont {Da~Costa}}, \bibinfo {author} {\bibfnamefont {S.}~\bibnamefont {Dorogovtsev}},\ and\ \bibinfo {author} {\bibfnamefont {J.}~\bibnamefont {Mendes}},\ }\href@noop {} {\bibfield  {journal} {\bibinfo  {journal} {Physical Review E}\ }\textbf {\bibinfo {volume} {94}},\ \bibinfo {pages} {062305} (\bibinfo {year} {2016})}\BibitemShut {NoStop}%
\bibitem [{\citenamefont {Pastor-Satorras}\ and\ \citenamefont {Castellano}(2018)}]{pastor2018eigenvector}%
  \BibitemOpen
  \bibfield  {author} {\bibinfo {author} {\bibfnamefont {R.}~\bibnamefont {Pastor-Satorras}}\ and\ \bibinfo {author} {\bibfnamefont {C.}~\bibnamefont {Castellano}},\ }\href@noop {} {\bibfield  {journal} {\bibinfo  {journal} {Journal of Statistical Physics}\ }\textbf {\bibinfo {volume} {173}},\ \bibinfo {pages} {1110} (\bibinfo {year} {2018})}\BibitemShut {NoStop}%
\bibitem [{\citenamefont {de~Arruda}\ \emph {et~al.}(2017)\citenamefont {de~Arruda}, \citenamefont {Cozzo}, \citenamefont {Peixoto}, \citenamefont {Rodrigues},\ and\ \citenamefont {Moreno}}]{de_Arruda_2017}%
  \BibitemOpen
  \bibfield  {author} {\bibinfo {author} {\bibfnamefont {G.~F.}\ \bibnamefont {de~Arruda}}, \bibinfo {author} {\bibfnamefont {E.}~\bibnamefont {Cozzo}}, \bibinfo {author} {\bibfnamefont {T.~P.}\ \bibnamefont {Peixoto}}, \bibinfo {author} {\bibfnamefont {F.~A.}\ \bibnamefont {Rodrigues}},\ and\ \bibinfo {author} {\bibfnamefont {Y.}~\bibnamefont {Moreno}},\ }\bibfield  {journal} {\bibinfo  {journal} {Physical Review X}\ }\textbf {\bibinfo {volume} {7}},\ \href {https://doi.org/10.1103/physrevx.7.011014} {10.1103/physrevx.7.011014} (\bibinfo {year} {2017})\BibitemShut {NoStop}%
\bibitem [{\citenamefont {Cozzo}\ \emph {et~al.}(2013)\citenamefont {Cozzo}, \citenamefont {Baños}, \citenamefont {Meloni},\ and\ \citenamefont {Moreno}}]{Cozzo_2013}%
  \BibitemOpen
  \bibfield  {author} {\bibinfo {author} {\bibfnamefont {E.}~\bibnamefont {Cozzo}}, \bibinfo {author} {\bibfnamefont {R.~A.}\ \bibnamefont {Baños}}, \bibinfo {author} {\bibfnamefont {S.}~\bibnamefont {Meloni}},\ and\ \bibinfo {author} {\bibfnamefont {Y.}~\bibnamefont {Moreno}},\ }\bibfield  {journal} {\bibinfo  {journal} {Physical Review E}\ }\textbf {\bibinfo {volume} {88}},\ \href {https://doi.org/10.1103/physreve.88.050801} {10.1103/physreve.88.050801} (\bibinfo {year} {2013})\BibitemShut {NoStop}%
\bibitem [{\citenamefont {Ferraz~de Arruda}\ \emph {et~al.}(2020)\citenamefont {Ferraz~de Arruda}, \citenamefont {Méndez-Bermúdez}, \citenamefont {Rodrigues},\ and\ \citenamefont {Moreno}}]{Ferraz_de_Arruda_2020}%
  \BibitemOpen
  \bibfield  {author} {\bibinfo {author} {\bibfnamefont {G.}~\bibnamefont {Ferraz~de Arruda}}, \bibinfo {author} {\bibfnamefont {J.~A.}\ \bibnamefont {Méndez-Bermúdez}}, \bibinfo {author} {\bibfnamefont {F.~A.}\ \bibnamefont {Rodrigues}},\ and\ \bibinfo {author} {\bibfnamefont {Y.}~\bibnamefont {Moreno}},\ }\href {https://doi.org/10.1088/1742-5468/abbcd4} {\bibfield  {journal} {\bibinfo  {journal} {Journal of Statistical Mechanics: Theory and Experiment}\ }\textbf {\bibinfo {volume} {2020}},\ \bibinfo {pages} {103405} (\bibinfo {year} {2020})}\BibitemShut {NoStop}%
\bibitem [{\citenamefont {Mauro}\ \emph {et~al.}(2009)\citenamefont {Mauro}, \citenamefont {Loucks},\ and\ \citenamefont {Gupta}}]{mauro2009fictive}%
  \BibitemOpen
  \bibfield  {author} {\bibinfo {author} {\bibfnamefont {J.~C.}\ \bibnamefont {Mauro}}, \bibinfo {author} {\bibfnamefont {R.~J.}\ \bibnamefont {Loucks}},\ and\ \bibinfo {author} {\bibfnamefont {P.~K.}\ \bibnamefont {Gupta}},\ }\href@noop {} {\bibfield  {journal} {\bibinfo  {journal} {Journal of the American Ceramic Society}\ }\textbf {\bibinfo {volume} {92}},\ \bibinfo {pages} {75} (\bibinfo {year} {2009})}\BibitemShut {NoStop}%
\bibitem [{\citenamefont {G{\'o}mez}\ \emph {et~al.}(2010)\citenamefont {G{\'o}mez}, \citenamefont {Arenas}, \citenamefont {Borge-Holthoefer}, \citenamefont {Meloni},\ and\ \citenamefont {Moreno}}]{gomez2010discrete}%
  \BibitemOpen
  \bibfield  {author} {\bibinfo {author} {\bibfnamefont {S.}~\bibnamefont {G{\'o}mez}}, \bibinfo {author} {\bibfnamefont {A.}~\bibnamefont {Arenas}}, \bibinfo {author} {\bibfnamefont {J.}~\bibnamefont {Borge-Holthoefer}}, \bibinfo {author} {\bibfnamefont {S.}~\bibnamefont {Meloni}},\ and\ \bibinfo {author} {\bibfnamefont {Y.}~\bibnamefont {Moreno}},\ }\href@noop {} {\bibfield  {journal} {\bibinfo  {journal} {Europhysics Letters}\ }\textbf {\bibinfo {volume} {89}},\ \bibinfo {pages} {38009} (\bibinfo {year} {2010})}\BibitemShut {NoStop}%
\bibitem [{\citenamefont {de~Oliveira}\ and\ \citenamefont {Dickman}(2004)}]{deoliveira2004simulate}%
  \BibitemOpen
  \bibfield  {author} {\bibinfo {author} {\bibfnamefont {M.~M.}\ \bibnamefont {de~Oliveira}}\ and\ \bibinfo {author} {\bibfnamefont {R.}~\bibnamefont {Dickman}},\ }\href@noop {} {\bibinfo {title} {How to simulate the quasi-stationary state}} (\bibinfo {year} {2004}),\ \Eprint {https://arxiv.org/abs/cond-mat/0407797} {arXiv:cond-mat/0407797 [cond-mat.stat-mech]} \BibitemShut {NoStop}%
\end{thebibliography}%

\end{document}